\documentclass[letterpaper, 10 pt, conference]{ieeeconf}
\IEEEoverridecommandlockouts
\pdfoutput=1

\usepackage{booktabs}
\usepackage{glossaries}
\usepackage{glossary-longbooktabs}
\usepackage{pgfplots}
\usepackage{multirow}
\usepackage{tablefootnote}
\pgfplotsset{compat=1.17}
\usepackage{xcolor}

\title{\LARGE \bf
Valuing Uncertainties in Wind Generation: An Agent-Based Optimization Approach
}

\author{Daniel Shen$^{1}$ and Marija Ilic$^{2}$
\thanks{$^{1}$Daniel Shen is a PhD student in the MIT Department of Electrical Engineering and Computer Science
        {\tt\small oski@mit.edu}}%
\thanks{$^{2}$Marija Ilic is a professor with the MIT Department of Electrical Engineering and Computer Science
        {\tt\small ilic@mit.edu}}%
}

\newglossary[flg]{skip}{fls}{flo}{Fake Entries}
\makeglossaries
\newglossaryentry{generator-dispatch-t}{
    name=\ensuremath{p_{i,t}},
    description={Dispatch of generator $i$ at time $t$},
    sort=1
}
\newglossaryentry{generator-dispatch-tminus}{
    name=\ensuremath{p_{i,t-1}},
    description={Dispatch of generator $i$ at time $t-1$},
    type=skip
}
\newglossaryentry{generator-limit-max}{
    name=\ensuremath{\overline{p}_{i,t}},
    description={Maximum dispatch of generator (MW) at time $t$},
    sort=3
}
\newglossaryentry{generator-limit-min}{
    name=\ensuremath{\underline{p}_{i,t}},
    description={Minimum dispatch of generator (MW) at time $t$},
    sort=2
}
\newglossaryentry{generator-quadratic}{
    name=\ensuremath{a_{i,t}},
    description={Quadratic coefficient of generator total offer curve},
    sort=4
}
\newglossaryentry{generator-linear}{
    name=\ensuremath{b_{i,t}},
    description={Linear coefficient of generator total offer curve},
    sort=5
}
\newglossaryentry{generator-constant}{
    name=\ensuremath{c_{i,t}},
    description={Constant coefficient of generator total offer curve},
    sort=6
}
\newglossaryentry{unit-commit}{
    name=\ensuremath{u_{i,t}},
    description={On/off status for generator with unit commitment},
    sort=7
}
\newglossaryentry{generator-ramp}{
    name=\ensuremath{RR_{i, t}},
    description={Ramp rate for conventional generators},
    sort=10
}
\newglossaryentry{flow-power}{
    name=\ensuremath{\emph{f}_{l,t}},
    description={Flow of power in branch $l$ at time $t$},
    sort=13
}
\newglossaryentry{phase-angle-diff}{
    name=\ensuremath{\delta \theta_{n,m,t}},
    description={Phase angle difference between buses $n, m$ at time $t$},
    type=skip
}

\newglossaryentry{phase-angle}{
    name=\ensuremath{\theta_{n,t}},
    description={Phase angle at bus $n$ at time $t$},
    sort=11
}
\newglossaryentry{line-reactance}{
    name=\ensuremath{x_l},
    description={Line reactance},
    sort=12
}
\newglossaryentry{branch-capacity}{
    name=\ensuremath{F_l},
    description={Capacity of branch l},
    sort=14
}
\newglossaryentry{startup-cost}{
    name=\ensuremath{suc_{i,t}},
    description={Startup cost if generator is started at time $t$},
    sort=8
}
\newglossaryentry{shutdown-cost}{
    name=\ensuremath{sdc_{i,t}},
    description={Shutdown cost if generator is shut down at time $t$},
    sort=9
}
\newglossaryentry{day-ahead-price}{
    name=\ensuremath{\lambda_{i,t}},
    description={Day-ahead marginal price for generator $i$ at $t$}
}
\newglossaryentry{real-time-price}{
    name=\ensuremath{\alpha_{i,t}},
    description={Real-time marginal price for generator $i$ at $t$}
}

\begin{document}

\maketitle
\thispagestyle{empty}
\pagestyle{empty}

\begin{abstract}

The increasing integration of variable renewable energy sources such as wind and solar will require new methods of managing generation uncertainty. Existing practices of uncertainty management for these resources largely focuses around modifying the energy offers of such resources in the quantity domain and from a centralized system operator consideration of these uncertainties. This paper proposes an approach to instead consider these uncertainties in the price domain, where more uncertain power is offered at a higher price instead of restricting the quantity offered. We demonstrate system-level impacts on a modified version of the RTS-GMLC system where wind generators create market offers valuing their uncertainties over scenario set of day-ahead production forecasts. The results are compared with a dispatch method in which wind energy is offered at zero marginal price and restricted based on the forecast percentile.

\end{abstract}

\printglossary[title=Nomenclature, nonumberlist, type=main]

\let\thefootnote\relax\footnote{The information, data, or work presented herein was supported by the
National Science Foundation Graduate Research Fellowship under Grant No. 2141064 as well as in part by the Advanced Research Projects Agency-Energy (ARPA-E), U.S. Department of Energy under Award Number DE-AR0001277. The views and opinions of authors expressed herein do not necessarily state or reflect those of the United States Government or any agency thereof.}

\section{INTRODUCTION}

Widespread integration of variable renewable energy (VRE) into the electrical grid is critical to reducing global carbon emissions. In contrast to conventional power plants, the capacity of VRE resources such as wind and solar power plants is subject to greater prediction uncertainties. As the proportion of VRE on the grid increases, managing these uncertainties and integrating their management into electricity markets and dispatch will become increasingly important. Managing generation uncertainty is not just an issue for renewables; in New York, the system operator has adopted air temperature-adjusted capacity ratings for internal combustion, combustion, and combined cycle units \cite{new_york_independent_system_operator_installed_2022}.

In regions of the USA with deregulated electricity markets, central system operators are responsible for the handling the bulk of market decisions and constraints. These responsibilities range from commitment decisions (which generators should turn on or off on any given day), procuring sufficient operating reserves to handle deviations in supply and demand, and ensuring that generators are dispatched with consideration to thier operating limits. Much of the research towards managing uncertainties in generator dispatch reflects this centralized paradigm. Some non-exhaustive examples of centralized generation uncertainty management include \cite{pena-ordieres_dc_2021, sundar_chance-constrained_2019, bukhsh_integrated_2016, zhang_distributionally_2016}, and \cite{bienstock_chance-constrained_2014}.

Several challenges come with centralized management of uncertainties. Requiring market participants to expose their internal constraints and data to the system operator comes with privacy concerns, especially when considering demand-side resources that involve individual customer data. In addition, scaling multiperiod optimization with generation ramp constraints, uncertainty management, and unit commitment decisions becomes computationally challenging due to the sheer number of decision variables; this problem will only continue to grow as an increasing number of smaller resources begin to participate in these markets. 

One alternative option to ameliorate these concerns is for resources to estimate the overall economic cost of their uncertainties and then communicate these costs to the system operator through price functions. This way, the responsibility for managing uncertainty shifts from the central system operator to individual market participants and the scope of the optimization problem at the system operators' level is reduced. Previous work using the framework of Dynamic Monitoring and Decision Systems (DYMONDS, see \cite{ilic_dynamic_2011} for more details) has demonstrated the ability to use such a framework for coordinating wind and price-responsive demands without exposing user privacy or increasing the number of decision variables at the system operator level \cite{ilic_efficient_2011}. A key part of the DYMONDS framework relevant to this work is the internalization of temporal uncertainties and risks at the resource level, this is enabled by interactive information exchange between stakeholders at various levels of the operations hierarchy to support distributed optimization.

In this paper, we consider how such a framework can be applied to valuing uncertainties in wind and solar generation forecasts and the effects of these valuations on system operations. We present methodology for ``uncertainty-aware'' offers based in individual agents' risk preferences. We compare this DYMONDS-based approach where wind and solar generation estimate the costs of their uncertainty with a centralized approach where a system operator dispatches VREs based on forecast heuristics. Results are demonstrated on the a modified RTS-GMLC system \cite{barrows_ieee_2019}, an updated version of the RTS-96 test system with a more contemporary generation mix and representative wind timeseries data.

\section{Deregulated Electricity Markets}
Although deregulated electricity markets vary in their specifics across the United States, there are common characteristics. In this section, we briefly outline how such markets operate and key points relevant to this work. For a more detailed overview of electricity markets, refer to the FERC Energy Primer \cite{federal_energy_regulatory_commission_energy_2022}.

\subsection{Energy Market Structure}
In deregulated markets, generators submit offer curves, ramping constraints, maximum and minimum power output limits, startup and shutdown costs, and minimum up/down information to the system operator. Based on this generator information and a load forecast, the system operator selects which generators will be run and at what output to meet the system load at the lowest cost while respecting factors such as transmission constraints and contingency consideration. This selection process occurs in two stages. 

In the day-ahead unit commitment (UC) stage, the system operator selects which generators will be turned on or off for the coming operating day to adequately meet the predicted load profile. At this stage, the system operator will also issue a dispatch schedule for each hour of the coming day specifying at what level each generator will run. 

The second stage, real-time dispatch (or system dispatch), involves changing generator dispatch levels in response to improved forecasts in load and grid conditions as the operating time approaches. This occurs approximately one hour before the operating hour. If additional generation capacity must be committed to meet the load or decommitted because too much capacity is online, fast-start units such as gas peaker plants may have their commitment decisions changed at this stage.

\subsection{Generator Offers}
Generators are expected to submit offer curves that are reflective of their variable costs (fuel and O\&M expenses). These curves reflect the marginal cost for providing electricity at various generator output levels. For thermal power plants, the heat rate and thus marginal cost vary as a function of power generated. For VRE resources, the marginal cost is typically submitted as a constant zero or even negative price across the generator's capacity for each interval, with the maximum power offered in each interval being checked against a centralized forecast for wind or solar capacity \cite{bird_wind_2014}. Submitting noncompetitive offers which are excess of a generator's actual variable costs can be penalized by market power regulators as an attempt to raise market prices.

\subsection{Prices and Payments}
Locational marginal prices (LMPs) or zonal marginal prices are calculated for each generator based on the offer curves, transmission constraints, and load conditions. Generators are paid for the amount of power dispatched at each stage multiplied by the LMP. Generators pay (or are paid) for differences between the MW dispatched in the day-ahead and real-time markets.

Generators which require commitment decisions will also have startup and shutdown costs associated with changes in generator status. When the revenue from LMP payments falls short of these costs, the system operator may issue ``uplift'' credits to these generators to cover the shortfall. A discussion of uplift payments is beyond the scope of this paper, which focuses on offer-based LMPs.

\section{METHODS}

We simulated four days of simplified energy market operations to illustrate the differences between ``centralized'' percentile-based VRE uncertainty management and DYMONDS-based uncertainty management where individual resources create uncertainty-aware offer curves. The system operator's optimization problem was formulated as a DC optimal power flow (DCOPF) with unit commitment considerations. The DCOPF was solved using a modified version of PyPSA \cite{PyPSA}.

\subsection{Unit Commitment}
This section summarizes the unit commitment formulations used in both sections \ref{section:central-dispatch} and \ref{section:dymonds-dispatch}. In our DCOPF formulation of the unit commitment problem, we aim to minimize the total payments to generators, subject to transmission, generator ramp, and capacity constraints. In addition, unlike most other formulations of the DCOPF in the literature, we allow the generator offer curves to vary as a function of time.

The overall optimization problem is given as

\begin{align}
    min \sum_{i,t} & \gls{generator-quadratic} \gls{generator-dispatch-t}^2 + \gls{generator-linear} \gls{generator-dispatch-t} + \gls{generator-constant} \gls{unit-commit} + \gls{startup-cost} + \gls{shutdown-cost} \label{eq:uc-objective} \\
    s.t. \quad &\gls{unit-commit} \gls{generator-limit-min} \leq \gls{generator-dispatch-t} \leq \gls{unit-commit} \gls{generator-limit-max} \label{eq:constraint-pmax}\\
    & |\gls{generator-dispatch-t} - \gls{generator-dispatch-tminus}| \leq \gls{generator-ramp} \label{eq:constraint-ramp} \\
    & \gls{flow-power} = \frac{\gls{phase-angle-diff}}{\gls{line-reactance}} \label{eq:constraint-phaseangle} \\
    & |\gls{flow-power}| \leq \gls{branch-capacity} \label{eq:constraint-linecapacity}
\end{align}

Where the variables are the generator dispatches \gls{generator-dispatch-t}, binary commitment decisions \gls{unit-commit}, and bus phase angles \gls{phase-angle}. 
\begin{itemize}
\item \gls{generator-quadratic}, \gls{generator-linear}, and \gls{generator-constant} represent the generation costs and must form a convex function.
\item (\ref{eq:constraint-pmax}) sets the generators' min \& max production, subject to the UC status.
\item (\ref{eq:constraint-ramp}) limits the maximum upwards and downwards ramp rate of the generator.
\item (\ref{eq:constraint-phaseangle}) specifies the power flow from the bus phase angle differences.
\item (\ref{eq:constraint-linecapacity}) restricts the maximum power flow on a line to that line's rated capacity.

\end{itemize}

In addition to constraints (\ref{eq:constraint-pmax}) - (\ref{eq:constraint-linecapacity}), a minimum up and down time are imposed for committable generators.

Since the dual variables (LMPs) can only be computed when the binary commitment decisions are fixed, the marginal prices must be generated by rerunning the problem once the UC decisions have been made.

In our simulations, the UC was run over a horizon of 26h in the day-ahead market and a rolling horizon of 3h in the real-time market.

\subsection{Centralized Percentile VRE Management and Dispatch}
\label{section:central-dispatch}

In the centralized percentile model, all generators make offers that reflect their variable fuel costs; these offer curves are thus constant as a function of time. In this model, wind generators have zero marginal costs. Uncertainties in wind production are managed by restricting the maximum quantity of wind (\gls{generator-limit-max}) offered in each interval of the day-ahead market at a given percentile of the scenario distribution for that hour.

In the real-time dispatch, the ``true'' quantity of wind that can be produced is revealed for each hour in the optimization horizon. If this quantity is lower than the amount that was dispatched in the day-ahead market, the wind generator is penalized for the shortfall at the real-time price. If instead the real-time quantity is higher than the day-ahead dispatch, then this surplus is reflected when the system operator reruns UC in real-time; since wind is offered at zero cost, a surplus of wind in the real-time market can displace conventional generation that was dispatched in the day-ahead market.

\subsection{Dispatch with Risk-Aware VRE Offers}
\label{section:dymonds-dispatch}

In the risk-aware DYMONDS model, offer curves are no longer restricted to reflecting variable costs. Instead of restricting offer curves to reflect wind generators' zero fuel costs, the wind generators calculate offer curves which reflect the penalties and risks associated with underdelivery of power in the real-time market. Since the specific penalty and risk varies from hour-to-hour as the forecast distribution, day-ahead prices, and real-time prices change, the optimal offer curve should also vary in time. 

\begin{figure}[!b]
    \begin{center}
        \scalebox{0.7}{\input{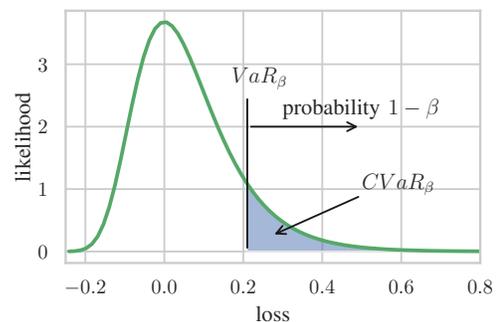}}
    \end{center}
    \vspace*{-5mm}
    \caption{Graphical representation of VaR and CVaR}
    \label{fig:cvar-explain}
\end{figure}

We use conditional value at risk (CVaR) \cite{rockafellar_optimization_2000} as a risk metric for constructing risk-aware offer curves since it is convex and the effects of different risk preferences can be observed by adjusting the $\beta$ parameter. CVaR measures the expected value of the tail losses that occur with cumulative probability $1 - \beta$ and is sensitive to the worst-case scenarios in the loss distribution. Fig. \ref{fig:cvar-explain} gives a graphical representation of conditional value at risk, the $\beta$ parameter, and value at risk (VaR).

As in the centralized model, any shortfall in day-ahead wind production must be repurchased at the real-time marginal price. The distribution of potential profits \& losses for a given generator are thus a function of what the distribution of wind scenarios, day-ahead prices, and real-time energy prices will be over a given operating day. 

Building upon the conditional value at risk framework described in \cite{yin_toward_2015}, variable generators create their risk-aware offers for each hour of the day-ahead market by solving the profit optimization problem:

\begin{align}
    \max_{a, b} & \quad  CVaR_{\beta, \gls{day-ahead-price}, \gls{real-time-price}} \left(\gls{day-ahead-price} \cdot \frac{\gls{day-ahead-price} - b}{2a} - \gls{real-time-price} \cdot \Delta \tilde{p}\right) \label{eq:cvar-optimization} \\
    s.t. & \quad a \geq 0 \label{constraint:dymonds-convex}
\end{align}

Here, $a$ and $b$ are the quadratic and linear coefficients of a generators' offer curve. Constraint \ref{constraint:dymonds-convex} ensures the marginal price is monotonically increasing to keep (\ref{eq:uc-objective}) convex. Assuming no congestion, under merit-order dispatch a generator will be dispatched for $ \frac{\gls{day-ahead-price} - b}{2a}$ MW of power.

The term $\Delta \tilde{p}$ represents the distribution of real-time shortfalls for a given day-ahead dispatch. Because the case where a generator intentionally reduces capacity in the day-ahead market in anticipation of offering it in the real-time market can be penalized by regulators as withholding, the variable $\Delta \tilde{p}$ is floored to be strictly positive.

Estimating the distributions of \gls{day-ahead-price} and \gls{real-time-price} is difficult, since their values are dependent on system conditions and other generators' offers and strategies. In this work, we estimate the distributions of \gls{day-ahead-price} and \gls{real-time-price} as follows:

\begin{enumerate}
    \item The hourly mean and covariance of day-ahead and real-time prices are calculated from centralized results when wind generators offer at maximum their median forecast in the day-ahead market. \label{day-ahead-mean-cov}
    \item Points from the distribution in step \ref{day-ahead-mean-cov} are sampled to create a scenario set of prices that are combined with the wind scenarios to yield an hourly profit/loss distribution upon which the CVaR optimization in Eq. \ref{eq:cvar-optimization} is applied. \label{step2}
    \item The system is dispatched using the risk-aware offer curves for each hour to generate the final results.
\end{enumerate}

Additionally, instead of setting $\gls{generator-limit-max}$ based on a percentile of the wind scenario distribution in each hour, the offer is capped by the highest production scenario in each hour.

\section{CASE STUDY}

Two consecutive days of operations (2020-07-17 to 2020-07-18) were simulated on a modified version of the RTS-GMLC system. In all simulations, the day-ahead and real-time markets were cleared at 1h intervals. Although most real-time markets operate on a higher resolution (5m - 15m), the scenario data for load and VRE generation is only provided at the 1h resolution.

The CVaR $\beta$ parameter was evaluated at 0.25, 0.5, and 0.75 for the wind generators to examine the effect of risk aversity on generator offers and overall system outcomes.

\subsection{Test System Overview}
The RTS-GMLC network is a modified version of the RTS-96 test system with an updated generation mix and wind and solar timeseries forecasts. We used PGScen \cite{carmona_joint_2022} to create 10 representative load and wind scenarios over the two operating days.

The RTS-GMLC system's generation capacity is significantly overbuilt for normal operating conditions (no contingency considerations) at its July peak of ~11.3 GW. We thus removed several generation categories from the network to better observe the effects of wind uncertainty under different optimization approaches. The generation mix of the modified RTS-GMLC system used in this work is given in Table \ref{table:rts-gmlc}.

\subsection{Risk-Aware VRE Offers}

We applied risk-aware offer creation (Section \ref{section:dymonds-dispatch}) for 3 different values of $\beta$; in each $\beta$ set, all wind generators had the same risk preference factor. A higher value of $\beta$ indicates a greater aversion to risk.

\begin{figure}[!h]
    \centering
    \setlength{\fboxrule}{0pt}
        \framebox{\scalebox{0.8}{\input{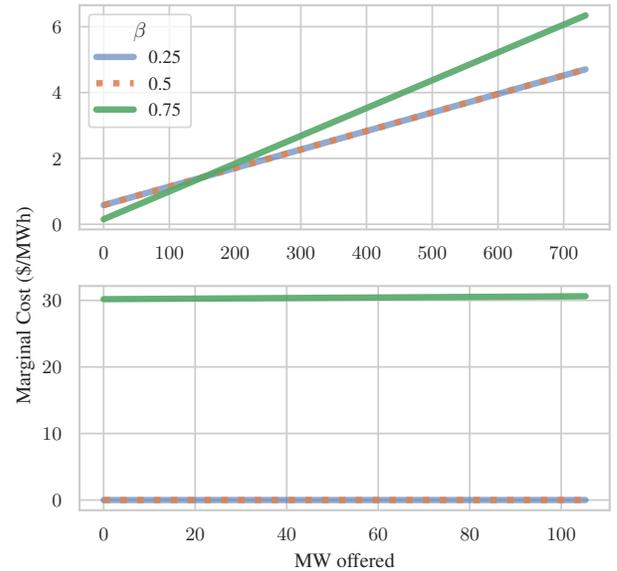}}}
    \caption{Wind generator offer curves for generator 317\_WIND\_1 at times 2020-07-17 12:00 (top) and 2020-07-17 16:00 (bottom)}
    \label{fig:dymonds-offers}
\end{figure}

Fig. \ref{fig:dymonds-offers} shows the difference between offer curve for a wind generator at different time intervals in the day-ahead market when Eq. \ref{eq:cvar-optimization} is applied. Both the maximum quantity offered and the slope of the marginal price change as the forecast scenario distribution changes. As the generator's risk aversion decreases (decreasing $\beta$), the slope of the marginal cost decreases. This reflects a greater willingness to take on potential losses in the real-time market; a less-steep marginal cost increases the chance that the generator will be dispatched at a higher capacity in the day-time market and may have to buy back a greater shortfall in real-time.

\begin{table}[t!]
\caption{RTS-GMLC Generation Capacities (MW) and Modifications}
\label{table:rts-gmlc}
\begin{center}
\resizebox{\columnwidth}{!}{
\begin{tabular}{@{}lrrrr@{}}
\toprule \\
\textbf{Generation}      & \textbf{\begin{tabular}[c]{@{}r@{}}{Total capacity} \\ {(RTS-GMLC)}\end{tabular}} & \textbf{Count} & \textbf{Keep?} & \textbf{Final \%} \\
\midrule \\
Gas (CC)                 & 3550                                                                         & 10             & Yes            & 31                               \\
Wind                     & 2508                                                                         & 4              & Yes            & 22                               \\
Coal                     & 2317                                                                       & 16             & Yes            & 20                               \\
Gas (CT)                 & 1485                                                                       & 27             & Yes            & 13                               \\
Hydro                    & 950                                                                        & 19             & Yes            & 8                                \\
Nuclear                  & 400                                                                        & 1              & Yes            & 3                                \\
Oil (CT)                 & 240                                                                        & 12             & Yes            & 2                                \\
Oil (Steam)              & 84                                                                         & 7              & Yes            & 1                                \\
\\
PV (Utility)             & 1555                                                                       & 25             & No             & 0                               \\
Rooftop PV               & 1161                                                                       & 31             & No             & 0                                \\
Concentrated Solar       & 200                                                                        & 1              & No             & 0                                \\
Run-of-River       & 50                                                                         & 1              & No             & 0                                \\
Storage                  & 50                                                                         & 1              & No             & 0                               \\
\bottomrule \\
\end{tabular}
}
\end{center}
\end{table}

\begin{figure}[!h]
    \centering
    \setlength{\fboxrule}{0pt}
        \framebox{\scalebox{0.8}{\input{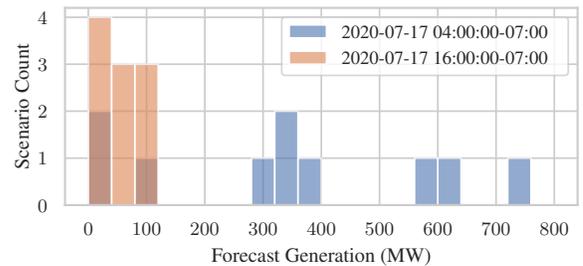}}}
    
    \caption{Comparison of wind scenario distributions for generator 317\_WIND\_1 at hours 0400 and 1600}
    \label{fig:forecast-compare}
\end{figure}

Furthermore, as the variance in generation scenarios increases over a single snapshot, the marginal cost curve is expected to steepen. A greater variance implies more uncertainty in the ability to deliver on MW at the higher end of the distribution. Dispatching the MW at the tail end is less desirable for both the system operator and the generator. The generator is more likely to incur the real-time penalty for these tail MW, while the system operator sees it as more risky to fill the demand requirements with non-deliverable power. Fig. \ref{fig:forecast-compare} shows two distributions for day-ahead forecasts for different hours. In hour 4, the forecast distribution is much more uncertain than hour 16, with a range of $\sim$750 MW across the generation scenarios. Referring back to Fig. \ref{fig:dymonds-offers}, we see the expected behavior that hour 4 (top plot) has a steeper marginal cost curve for the same risk preference when compared to the generator's offer in hour 16 (bottom). 

There is overlap between the curves for $\beta=0.5$ and $\beta=0.75$ in Fig. \ref{fig:dymonds-offers} snapshots; this likely a numerical issue requiring tuning the termination parameters of the optimizer.

\subsection{System Impacts}

This section compares overall system impacts as measured by resdispatch required between the real-time and day-ahead markets and overall payments to generators. 

The total payment to a generator for providing energy is based on uniform marginal pricing and given by:
\begin{equation}
    R = \lambda_{_{DA}} p_{_{DA}} + \lambda_{_{RT}} (p_{_{RT}} - p_{_{DA}})
    \label{eq:settlement-rules}
\end{equation}

Where 

\begin{itemize}
    \item $\lambda_{_{DA}}$ is the day-ahead locational marginal price
    \item $\lambda_{_{RT}}$ is the real-time locational marginal price
    \item $p_{_{DA}}$ is the power dispatched in the day-ahead market
    \item $p_{_{RT}}$ is the power dispatched in the real-time market
\end{itemize}

Note that all generators (regardless of category) pay a penalty if their real-time dispatch is lower than the day-ahead dispatch.

Table \ref{table:redispatch-difference} shows the difference in energy that is settled in the real-time market vs. the day-ahead market for each generator category across both simulation days. A negative entry indicates that a greater amount of energy was dispatched for that generation category in the day-ahead market than the real-time market (in other words, the generators were overall downward dispatched in real-time). In the centralized dispatch method, wind offers are restricted to the n$^\text{th}$ forecast percentile in the day-ahead market and subsequently to the scenario realization amount in the real-time market. The probability of wind underdelivery increases as the percentile increases, as reflected by the change in redispatch difference from 350 to -389 GWh between the 25th and 75th percentile.

This is exactly opposite from the $\beta$ parameter in the risk-aware wind offers, where a higher $\beta$ value represents a more risk-averse bid. It is important to note here there is no a one-to-one correspondence between the percentile and $\beta$ values. We see that as the $\beta$ parameter is adjusted upwards, the amount of wind that is downward dispatched decreases from -1013 GWh to -449 GWh, as expected by an increasing amount of risk aversion on the part of the generator. When the wind offer curves are steeper, the prices on the tail end of the curve are noncompetitive when compared with the generation costs of conventional generators, thus limiting the amount of wind that is dispatched in day-ahead through price rather than hard quantity restrictions.

Tables \ref{table:total-payments} and \ref{table:total-payments-std} show statistics reflecting total payments to generators from the settlement rules described in Eq. (\ref{eq:settlement-rules}). Under the percentile-based restrictions, all generators receive strictly positive payments (Table \ref{table:total-payments}). This is not necessarily the case when wind generators create risk-aware offers; it is possible to create offers that are consistently too risky and incur financial losses. There is another outcome to submitting more risk-aware offers: the standard deviation of profits can be significantly lowered at the tradeoff of expected profits; compare for example a standard deviation in profits over two days of \$2.3M ($\beta=0.75$) vs. \$7.6M (25$^{\text{th}}$ percentile) for wind generators. This may be a significant consideration for generators that value consistent profits and avoiding catastrophic financial penalties for high impact, low probability events.

\begin{table}[b!]
\caption{Mean redispatch differences (Gigawatt hours, real-time - day-ahead) by generation type and wind offer strategy}
\label{table:redispatch-difference}
\resizebox{\columnwidth}{!}{
\begin{tabular}{lrrrrrr}
\toprule \\
\multicolumn{1}{c}{\multirow{2}{*}{\textbf{Generators}}} & \multicolumn{3}{c}{\textbf{Centralized Dispatch}}                              & \multicolumn{3}{c}{\textbf{Risk-Aware Wind Offers}}                                  \\
\multicolumn{1}{c}{}                                     & \multicolumn{1}{c}{25th$^*$} & \multicolumn{1}{c}{50th} & \multicolumn{1}{c}{75th} & \multicolumn{1}{c}{$\beta$=0.25} & \multicolumn{1}{c}{0.5} & \multicolumn{1}{c}{0.75} \\
\midrule \\
\textbf{Coal}      & -1564                    & -1315                    & -783                     & 38                            & -396                    & -752                     \\
\textbf{Hydro}     & -1342                    & -1223                    & -752                     & -769                          & -804                    & -1000                    \\
\textbf{Gas (CC)}           & -1615                    & -1521                    & -1225                    & -372                          & -794                    & -1136                    \\
\textbf{Gas (CT)}           & 6668                     & 7583                     & 8463                     & 8956                          & 9076                    & 8877                     \\
\textbf{Nuclear} & 0                        & 0                        & 0                        & 0                             & 0                       & 0                        \\
\textbf{Oil (CT)}          & -12                      & -11                      & -20                      & 0                             & -21                     & 0                        \\
\textbf{Oil (Steam)}       & 0                        & 0                        & 0                        & 0                             & 0                       & 0                        \\
\textbf{Wind}       & 350                      & 89                       & -389                     & -1013                         & -739                    & -449                     \\
\bottomrule \\
\multicolumn{7}{l}{\scriptsize *25th percentile for limiting day-ahead dispatch of wind} \\
\end{tabular}

}
\end{table}

\begin{table}[b!]
\caption{Mean total payments to generators (\$, thousands) by generation type and wind offer strategy}
\label{table:total-payments}
\resizebox{\columnwidth}{!}{
\begin{tabular}{lrrrrrr}
\toprule \\
\multicolumn{1}{c}{\multirow{2}{*}{\textbf{Generators}}} & \multicolumn{3}{c}{\textbf{Centralized Dispatch}}                              & \multicolumn{3}{c}{\textbf{Risk-Aware Wind Bids}}                                  \\
\multicolumn{1}{c}{}                                     & \multicolumn{1}{c}{25th} & \multicolumn{1}{c}{50th} & \multicolumn{1}{c}{75th} & \multicolumn{1}{c}{$\beta$=0.25} & \multicolumn{1}{c}{0.5} & \multicolumn{1}{c}{0.75} \\
\midrule
\textbf{Coal}                                      & 197432                   & 195418                   & 195407                   & 189988                        & 217076                  & 207836                   \\
\textbf{Hydro}                                     & 91328                    & 89724                    & 88916                    & 88494                         & 101349                  & 99549                    \\
\textbf{Gas (CC)}                                           & 236498                   & 234349                   & 224971                   & 205979                        & 238631                  & 241216                   \\
\textbf{Gas (CT)}                                           & 72343                    & 70984                    & 73424                    & 77706                         & 79749                   & 76999                    \\
\textbf{Nuclear}                                 & 45687                    & 45294                    & 44749                    & 41594                         & 47462                   & 46026                    \\
\textbf{Oil (CT)}                                          & 49                       & 44                       & 70                       & 0                             & 1005                    & 0                        \\
\textbf{Oil (Steam)}                                       & 183                      & 182                      & 182                      & 183                           & 183                     & 185                      \\
\textbf{Wind}                                       & 32312                    & 35289                    & 24425                    & -9647                         & -610                    & 10240                    \\
 \\
\textbf{Total}                                           & 675832                   & 671284                   & 652144                   & 594297                        & 684845                  & 682051                  \\
\bottomrule \\
\end{tabular}
}
\end{table}

\begin{table}[bt!]
\centering
\caption{Standard deviation of total payments to generators (\$, thousands) by generation type and wind offer strategy}
\label{table:total-payments-std}
\begin{tabular}{lrrrrrr}
\toprule
\multicolumn{1}{c}{\multirow{2}{*}{\textbf{Generators}}} & \multicolumn{3}{c}{\textbf{Centralized Dispatch}}                              & \multicolumn{3}{c}{\textbf{Risk-Aware Wind Bids}}                                  \\
\multicolumn{1}{c}{}                                     & \multicolumn{1}{c}{25th} & \multicolumn{1}{c}{50th} & \multicolumn{1}{c}{75th} & \multicolumn{1}{c}{$\beta$=0.25} & \multicolumn{1}{c}{0.5} & \multicolumn{1}{c}{0.75} \\
\midrule
\textbf{Coal}                                      & 4146                     & 4180                     & 4320                     & 3886                          & 2825                    & 1086                     \\
\textbf{Hydro}                                     & 2291                     & 1954                     & 1588                     & 858                           & 816                     & 840                      \\
\textbf{NG (CC)}                                           & 3036                     & 3077                     & 3429                     & 3431                          & 2887                    & 1528                     \\
\textbf{NG (CT)}                                           & 2404                     & 2624                     & 3257                     & 3089                          & 2671                    & 1250                     \\
\textbf{Nuclear}                                 & 14                       & 13                       & 14                       & 14                            & 9                       & 4                        \\
\textbf{Oil (CT)}                                          & 11                       & 8                        & 12                       & 0                             & 6                       & 0                        \\
\textbf{Oil (Steam)}                                       & 0                        & 0                        & 0                        & 0                             & 0                       & 0                        \\
\textbf{Wind}                                       & 7572                     & 8160                     & 9049                     & 9138                          & 6582                    & 2331                    \\
\bottomrule
\end{tabular}
\end{table}

In both the $\beta=0.25$ and $\beta=0.5$ risk choices, wind generators overall receive \textit{negative profits} of -\$9.6M and \mbox{-\$610K}, respectively, and must pay money into the market. This is likely due to the crude estimation procedure for the distribution of day-ahead prices and recourse costs $\gls{day-ahead-price}$ and $\gls{real-time-price}$ that are used in Eq. \ref{eq:cvar-optimization}. Unlike in the centralized dispatch method, where wind generators bid zero marginal cost and (in the absence of congestion) are guaranteed to have at least a portion of their offer accepted, there is no such guarantee in the risk-aware offer approach when uncertainty is priced in. Miscalculating the optimal curve from a inaccurate assessment of the price and recourse distributions creates possibilities of incurring significant financial penalties, albeit at the benefit to the consumer of lower overall generation costs. Note that additionally, it is possible to be so risk-averse that even with a poor estimation of these distributions, on average wind generators can still make a profit, albeit smaller than what they would have accumulated with strictly following a percentile-based dispatch rule (\$10M for $\beta=0.75$ vs. \$32M, \$35M, and \$24M for percentile values of 25, 50, and 75).

It is almost guaranteed that the distribution estimation procedure described in Section \ref{section:dymonds-dispatch} is not representative of the true underlying distribution since it is based the percentile approach, wherein there is a significant chance for a generator to underoffer in the day-ahead market and then sell additional excess in the real-time market. This is in contrast with the risk-valuation based approach, which implies overoffering in the day-ahead (\gls{generator-limit-max} set to the maximum scenario) and then buying-back in real time. More accurate estimation of these distributions (e.g. through statistical learning) would undoubtedly improve the mean revenue of wind generators and is a subject for future work.

\section{CONCLUDING REMARKS}

This paper presents a method for variable renewable generators to create offer curves that reflect the financial risks of generation uncertainty in the day-ahead market by using the conditional value at risk (CVaR). We demonstrate the impacts of wind generators adopting a CVaR strategy on a test system with $22\%$ wind penetration (Table \ref{table:rts-gmlc}) and show that such a strategy has mixed effects on the revenues of wind generators and overall generator payments. The volatility of revenue for wind generators can be reduced at the expense of smaller expected profits. System-wide, generator revenues as a whole can decrease in volatility by 50-75\% with only a small (1\%) increase in overall generator payments.

The CVaR offer strategy requires the assumption of a day-ahead and real-time price distribution. Such distributions are difficult to obtain and justify for a test system where only a few days of operation under very different assumptions are simulated. Exploring the robustness of these results to differences in these price distributions is a possible future direction, as well as improved estimation of the underlying distributions through longer simulations and application of statistical learning.

Current prevailing discussion of generation participation in deregulated electricity markets assumes generators make offers reflecting marginal fuel costs. Management of uncertainty in the price domain, instead of the quantity domain as is currently the norm, would necessitate a change in these market procedures and market monitoring practices to more closely delineate what is considered as uncertainty management vs. ``gaming'' and exercise of market power. Future work will continue to explore the utility and implications of expressing resource-level constraints and uncertainty considerations in the price domain and its implications and interplay with existing market structures and stakeholders.

Finally, this work focuses on the management of financial uncertainties in the short term electricity markets. While appropriate management of financial uncertainties can provide some price signals for longer term reliability and planning, a full consideration of grid reliability requires more than just analyzing generators as financial agents. In this regard, the role of additional market mechanisms and institutions (e.g. the system operator's role in managing capacity markets) must be considered as well; the authors welcome further work and discussion on the topic.

\bibliographystyle{IEEEtran}
\bibliography{IEEEabrv,biblio.bib}

\begin{thebibliography}{10}
\providecommand{\url}[1]{#1}
\csname url@samestyle\endcsname
\providecommand{\newblock}{\relax}
\providecommand{\bibinfo}[2]{#2}
\providecommand{\BIBentrySTDinterwordspacing}{\spaceskip=0pt\relax}
\providecommand{\BIBentryALTinterwordstretchfactor}{4}
\providecommand{\BIBentryALTinterwordspacing}{\spaceskip=\fontdimen2\font plus
\BIBentryALTinterwordstretchfactor\fontdimen3\font minus
  \fontdimen4\font\relax}
\providecommand{\BIBforeignlanguage}[2]{{%
\expandafter\ifx\csname l@#1\endcsname\relax
\typeout{** WARNING: IEEEtran.bst: No hyphenation pattern has been}%
\typeout{** loaded for the language `#1'. Using the pattern for}%
\typeout{** the default language instead.}%
\else
\language=\csname l@#1\endcsname
\fi
#2}}
\providecommand{\BIBdecl}{\relax}
\BIBdecl

\bibitem{new_york_independent_system_operator_installed_2022}
{New York Independent System Operator}, ``Installed {Capacity} {Manual}
  {Attachments},'' Jun. 2022.

\bibitem{pena-ordieres_dc_2021}
A.~Peña-Ordieres, D.~K. Molzahn, L.~A. Roald, and A.~Wächter, ``{DC}
  {Optimal} {Power} {Flow} {With} {Joint} {Chance} {Constraints},'' \emph{IEEE
  Transactions on Power Systems}, vol.~36, no.~1, pp. 147--158, Jan. 2021.

\bibitem{sundar_chance-constrained_2019}
K.~Sundar, H.~Nagarajan, L.~Roald, S.~Misra, R.~Bent, and D.~Bienstock,
  ``Chance-{Constrained} {Unit} {Commitment} {With} {N}-1 {Security} and {Wind}
  {Uncertainty},'' \emph{IEEE Transactions on Control of Network Systems},
  vol.~6, no.~3, pp. 1062--1074, Sep. 2019.

\bibitem{bukhsh_integrated_2016}
W.~A. Bukhsh, C.~Zhang, and P.~Pinson, ``An {Integrated} {Multiperiod} {OPF}
  {Model} {With} {Demand} {Response} and {Renewable} {Generation}
  {Uncertainty},'' \emph{IEEE Transactions on Smart Grid}, vol.~7, no.~3, pp.
  1495--1503, May 2016.

\bibitem{zhang_distributionally_2016}
Y.~Zhang, S.~Shen, and J.~Mathieu, ``\BIBforeignlanguage{en}{Distributionally
  {Robust} {Chance}-{Constrained} {Optimal} {Power} {Flow} with {Uncertain}
  {Renewables} and {Uncertain} {Reserves} {Provided} by {Loads}},''
  \emph{\BIBforeignlanguage{en}{IEEE Transactions on Power Systems}}, pp. 1--1,
  2016.

\bibitem{bienstock_chance-constrained_2014}
D.~Bienstock, M.~Chertkov, and S.~Harnett,
  ``\BIBforeignlanguage{en}{Chance-{Constrained} {Optimal} {Power} {Flow}:
  {Risk}-{Aware} {Network} {Control} under {Uncertainty}},''
  \emph{\BIBforeignlanguage{en}{SIAM Review}}, vol.~56, no.~3, pp. 461--495,
  Jan. 2014.

\bibitem{ilic_dynamic_2011}
M.~D. Ilić, ``Dynamic {Monitoring} and {Decision} {Systems} for {Enabling}
  {Sustainable} {Energy} {Services},'' \emph{Proceedings of the IEEE}, vol.~99,
  no.~1, pp. 58--79, Jan. 2011.

\bibitem{ilic_efficient_2011}
M.~D. Ilic, L.~Xie, and J.-Y. Joo, ``Efficient {Coordination} of {Wind} {Power}
  and {Price}-{Responsive} {Demand}—{Part} {I}: {Theoretical}
  {Foundations},'' \emph{IEEE Transactions on Power Systems}, vol.~26, no.~4,
  pp. 1875--1884, Nov. 2011.

\bibitem{barrows_ieee_2019}
C.~Barrows \emph{et~al.}, ``\BIBforeignlanguage{English}{The {IEEE}
  {Reliability} {Test} {System}: {A} {Proposed} 2019 {Update}},''
  \emph{\BIBforeignlanguage{English}{IEEE Transactions on Power Systems}},
  vol.~35, no.~1, Jul. 2019.

\bibitem{federal_energy_regulatory_commission_energy_2022}
{Federal Energy Regulatory Commission}, ``Energy {Primer} : {A} {Handbook} of
  {Energy} {Market} {Basics},'' p. 150, Apr. 2022.

\bibitem{bird_wind_2014}
L.~Bird, J.~Cochran, and X.~Wang, ``\BIBforeignlanguage{en}{Wind and {Solar}
  {Energy} {Curtailment}: {Experience} and {Practices} in the {United}
  {States}},'' Tech. Rep. NREL/TP--6A20-60983, 1126842, Mar. 2014.

\bibitem{PyPSA}
T.~Brown, J.~H\"orsch, and D.~Schlachtberger, ``{PyPSA: Python for Power System
  Analysis},'' \emph{Journal of Open Research Software}, vol.~6, no.~4, 2018.

\bibitem{rockafellar_optimization_2000}
R.~T. Rockafellar and S.~Uryasev, ``\BIBforeignlanguage{en}{Optimization of
  conditional value-at-risk},'' \emph{\BIBforeignlanguage{en}{The Journal of
  Risk}}, vol.~2, no.~3, pp. 21--41, 2000.

\bibitem{yin_toward_2015}
X.~Yin, M.~D. Ilić, and B.~Sinopoli, ``Toward design of risk-based real-time
  dispatch at value,'' in \emph{2015 {IEEE} {Power} {Energy} {Society}
  {Innovative} {Smart} {Grid} {Technologies} {Conference} ({ISGT})}, Feb. 2015,
  pp. 1--5.

\bibitem{carmona_joint_2022}
R.~Carmona and X.~Yang, ``Joint {Stochastic} {Model} for {Electric} {Load},
  {Solar} and {Wind} {Power} at {Asset} {Level} and {Monte} {Carlo} {Scenario}
  {Generation},'' Sep. 2022, preprint.

\end{thebibliography}

\clearpage

\end{document}